\documentclass[12pt]{article}
\usepackage{epsfig}\parskip 5pt plus 1pt
\newcommand{\be}{\begin{equation}}
\newcommand{\ee}{\end{equation}}
\newcommand{\bea}{\begin{eqnarray}}
\newcommand{\eea}{\end{eqnarray}}

\def\simge{\mathrel{%
   \rlap{\raise 0.511ex \hbox{$>$}}{\lower 0.511ex \hbox{$\sim$}}}}
\def\simle{\mathrel{
   \rlap{\raise 0.511ex \hbox{$<$}}{\lower 0.511ex \hbox{$\sim$}}}}

\begin{document}
\thispagestyle{empty}
\vspace*{1cm}
\begin{center}
{\Large{\bf Enabling pulse compression and proton acceleration in a modular
ICF driver for nuclear and particle physics applications} }\\
\vspace{.5cm}
F. Terranova$^{\rm a,}$,
S.V. Bulanov$^{\rm b,c}$, J.L. Collier$^{\rm d}$,
H. Kiriyama$^{\rm b}$, F. Pegoraro$^{\rm e}$ \\
\vspace*{1cm}
$^{\rm a}$ I.N.F.N.,  Laboratori Nazionali di Frascati,
Frascati (Rome), Italy \\
$^{\rm b}$ Advanced Photon Research Centre, JAERI, Kizu-cho, Kyoto-fu,  Japan \\
$^{\rm c}$ A.~M.~Prokhorov General Physics Institute of RAS, Moscow, Russia \\
$^{\rm d}$ Central Laser Facility, Rutherford Appleton Laboratory, Didcot, UK \\
$^{\rm e}$ Dip. di Fisica, Univ. di Pisa and CNISM, Pisa, Italy \\
\end{center}

\vspace{.3cm}
\begin{abstract}
\noindent
The existence of efficient ion acceleration regimes in collective
laser-plasma interactions opens up the possibility to develop
high-energy physics facilities in conjunction with projects for
inertial confinement nuclear fusion (ICF) and neutron spallation
sources.  In this paper, we show that the pulse compression requests
to make operative these acceleration mechanisms do not fall in
contradiction with current technologies for high repetition rate ICF
drivers. In particular, we discuss explicitly a solution that exploits
optical parametric chirped pulse amplification and the intrinsic
modularity of the lasers aimed at ICF.
\end{abstract}

\vspace*{\stretch{2}}
\begin{flushleft}
  \vskip 2cm
{ PACS: 41.75.Jv, 42.65.Yj, 52.38.Kd \\ } 
\end{flushleft}

\newpage

\section{Introduction}
\label{introduction}

Particle acceleration through collective short wavelength
electromagnetic effects has been pursued for
decades~\cite{ref:veksler}. The advent of laser has made it possible
to use the interaction of the laser light with the charged
particles~\cite{ref:shimoda} and the plasma~\cite{ref:tajima} and this
technique is still considered a viable alternative to traditional
RF-based boosters. Such a confidence has been strengthened after the
advent of wideband oscillators based on Ti:sapphire and the revolution
in power ultrafast lasers due to the development of Chirped Pulse
Amplification~\cite{ref:CPA} (CPA).  Production of energetic ions and
electrons has been reported by several experimental groups and new
breakthroughs are expected after the commissioning of the next
generation of multi-petawatt lasers.  Nonetheless, there is large
consensus on the fact that a well-understood and stable regime of
acceleration hasn't been achieved, yet. The lack of a reference
mechanism is the root of the best known drawbacks of laser
acceleration: strong dependence on the initial conditions, large
energy spread, poor light-to-particle energy conversion efficiency and
significant shot to shot variations.  These limitations, however, are
unlikely to be intrinsic features of laser acceleration.  This is
particularly clear for ion acceleration.  It has long been understood
that fast ion generation is related to the presence of hot
electrons~\cite{ref:Gurevich-Pitaevskii}. A variety of effects occur
when the main source of acceleration is charge displacement due to the
electron motion in the plasma or inductive electric fields from self
generated magnetic fields~\cite{ref:maksimchuk}. However, for higher
laser intensities ($10^{23} \ \mathrm{W/cm}^2$) acceleration results
from the laser pressure exerted to the comoving electron-ion system;
the latter acts as a progressively more opaque screen for the laser
light and, in this case, charge displacement only plays the role of
rectifier for the transversal laser field and as a medium for
electron-ion energy transfer during light absorption. This mechanism
is described in details in
\cite{ref:esirkepov,ref:NIM_neutrino,ref:RMP}.  It provides energy
transfer efficiency comparable to RF-based synchrotrons or even
cyclotrons and, more importantly, decouples the final ion energy from
the accelerated ion current (see Sec.\ref{sec:protons}). If this
mechanism were confirmed experimentally, it could represent the first
serious alternative to synchrotrons suited for high energy physics
(HEP) applications. As noted in \cite{ref:NIM_neutrino}, a proof of
principle of this radiation-pressure dominated (RPD) acceleration
mechanism is at the borderline of current technology, but the
possibility of using this technique to overcome the limitations of
traditional proton accelerators faces many additional difficulties.
In particular all present high power lasers operate at very low
repetition rate. This is a classical problem e.g. in inertial
confinement fusion (ICF), where the basic principle could be
demonstrated, for instance, at the National Ignition Facility (NIF) in
US~\cite{ref:NIF} on a single-shot basis. However, the ultimate use of
ICF to produce electric power will require repetition rates of the
order of tens of Hz, an increase of several order of magnitude
compared to the shot rate achievable with state-of-the-art fusion
laser technology. This rate cannot be achieved with flashlamp-pumped
neodymium-doped glass lasers that requires a significant interpulse
cooling time. There is, currently, a very large effort to find
alternative solutions and promising setups based on excimer lasers or
diode-pumped solid state lasers have been identified, offering, in
principle, the repetition rates needed for ICF. In
\cite{ref:NIM_neutrino}, we noted that the solution of the problem of
thermal stability and the exploitation of the RPD mechanism would open
up the fascinating possibility of a multipurpose driver for ICF and
HEP applications. However, we did not address explicitly the question
whether the power requirements for ICF could fall in contrast with the
pulse compression requests needed for the RPD acceleration to be fully
operative. More precisely, the strong constraints on the choice of the
amplifying medium (and therefore on the gain bandwidth) could make
impossible an appropriate pulse compression, so that the intensity
needed to operate in the RPD regime would not be reached by a
multi-pulse device.  In this paper we show that there exists at least
one particular configuration that is able to fulfill simultaneously
the two requirements; i.e. it offers a high repetition rate device
that can be implemented to design a multi-GeV proton booster with a
technology suited for ICF energy production.  This solution is based
on the exploitation of optical parametric chirped pulse
amplification~\cite{ref:dubietis,ref:ross} (OPCPA) and the intrinsic
modularity of ICF drivers\footnote{These drivers are constituted by an
ensemble of independent beamlines. Synchronous operation of the beams
is requested for fusion applications since the power must be delivered
in a single shot (maximum peak power) and illuminate uniformly the
fuel target. However, in most of the applications related to particle
and nuclear physics, the performance of the facility mainly depends on
the achieved mean power and asynchronous operation modes can be
envisaged. This simplifies substantially the technological challenge
of the feedback system for light synchronization.}  (see
Sec.\ref{sec:drivers}); it is described in details in
Sec.~\ref{sec:OPCPA}, while its potentiality as a new generation
proton driver is discussed in Sec.~\ref{sec:protons}

\section{Drivers for Inertial Confinement Fusion}
\label{sec:drivers}

Large scale flashlamp-pumped solid state lasers built for ICF are
inherently single shot devices, requiring several hours to recover
from thermal distortions. They are aimed at a proof-of-principle for
ICF but their scaling to a cost effective fusion reactor is highly non
trivial. In the last decade, three main research lines have been
investigated and much more efforts will be put in the forecoming years
if the proof-of-principle programs for ICF at NIF or at Laser
Megajoule (LMJ) in France is successful.  The first one exploits
traditional RF-based technologies for ion acceleration to transfer
energy to the target, trigger ignition and sustain burning.  This
approach profits of the enormous experience gained in particle
accelerators since the 50's and the large efficiency ($\sim 30-35$\%)
obtained at HEP facilities~\cite{ref:bieri} but, as a matter of fact,
the mean intensities and the required uniformity of target
illumination are well beyond current technology. The second and third
ones exploit lasers to ignite and sustain fusion and are aimed at
developing systems with much higher thermal yield than Nd:glass. They
are based on diode-pumped solid state lasers (DPSSL) or high power
excimer lasers.

\subsection{Diode-pumped solid state lasers}
The possibility of building ICF lasers with high repetition rates and
efficiency using solid state materials mainly relies on the
substitution of flashlamps with low-cost laser diode arrays and the
development of crystals for greater energy storage and thermal
conductivity than Nd:glass~\cite{ref:krupke}. Yb:crystals cooled by
near sonic helium jets are presently favorite candidates. The main
advantage of this approach is that it retains most of the features of
Nd:glass systems, first of all the possibility (still to be
demonstrated) of $<1\%$ smooth irradiation on-target for direct drive
in a timescale of fractions of ns. A DPSSL-based fusion reactor would
be - like NIF - highly modular. One possible vision is based on 4 kJ
DPSSL composed of 1 kJ beamlets operating at a repetition rate of the
order of 10~Hz and assembled to reach the overall MJ power per
shot. To our knowledge, the most advanced R\&D project is the
Mercury/Venus laser system~\cite{ref:pres_mercury}. In particular, the
Mercury R\&D is aimed at a 100 J, 10 Hz laser based on gas cooled
Yb:S-FAP crystals grown up to 20~cm diameter~\cite{ref:bibeau}. The
laser operates at 1047~nm (1$\omega$) with a 2~ns pulsewidth, a 5x
diffraction limited beam quality and an efficiency greater than
5\%. Fusion drivers are better operated at higher frequencies to
increase the rocket efficiency and reduce laser-plasma
instabilities~\cite{ref:sethian}. Hence, DPSSL are operated at
3$\omega$ (349~nm) with a conversion efficiency greater than 80\%.
Gain bandwidth is of the order of 5~nm for Yb:S-FAP, significantly
lower than for Nd:glass (28~nm) so that the time duration of a DPSSL
Fourier-limited (chirp-free) TEM$_{00}$ output beam would be bandwidth
limited to $\sim0.3$~ps pulses.

\subsection{Excimer lasers}
Excimer power lasers have been developed both for laser fusion and
defense uses. The current main candidate for ICF is
krypton-fluoride. Electron beam pumped KrF systems offer superior beam
spatial uniformity, short wavelength and high laser efficiency ($\sim
10\%$). As for DPSSL, an excimer-based fusion reactor is highly
modular and single beamlines could provide up to 50 kJ of laser
light~\cite{ref:sethian}. Again, the thermal yield and the
efficiencies requested for a viable commercial power
plant~\cite{ref:sombrero} represent major technological
challenges. The laser operates at 248 nm but a certain degree of
tunability is offered by the fact that the same system design can be
re-used for other gas mixtures~\cite{ref:smiley} (e.g. ArXe lasing at
1733 nm or XeF at 351 nm). In particular, XeF has been the leading
candidate for defense applications and large aperture lasers with
energy yield per pulse in the 5 kJ range has been built since the late
80's~\cite{ref:ewing}.  XeF has also been considered for laser fusion
but it is less effective than KrF due to its lower efficiency and
because it behaves spectrally inhomogeneous, precluding efficient
narrow-band operation~\cite{ref:ewing_1979}.

\section{OPCPA pulse compression}
\label{sec:OPCPA}

As a by-product of its peculiar design (see Sec.\ref{sec:drivers}), a
multi-shot ICF driver offers a large number of beamlines operating,
probably, in the near-UV region with a rather limited spectral
bandwidth and an energy per pulse ranging from 1 to 50 kJ. We do not
expect a Ti:sapphire CPA system being able to use efficiently neither
this pump source nor its outstanding average power regime. On the
other hand, Optical Parametric Chirped Pulse Amplification offers, in
principle, a higher degree of tunability and could be successfully
adapted to exploit a narrow band, energetic pump
pulse~\cite{ref:ross_las_part_beams} and its average
power~\cite{ref:velsko}. Pulse compression should be enough to trigger
the RPD acceleration mechanism and exploit the high repetition rate to
increase the average ion current\footnote{We do not intend this
technique as a route for Fast Ignition since the spectrum of the ions
is too energetic once the RPD regime is operational.}.
Optical parametric amplification is a nonlinear process that involves a
signal wave, a pump and an idler wave~\cite{ref:boyd}. In a suitable
nonlinear crystal, the high intensity and high frequency pump beam
($\omega_p$) amplifies a lower frequency ($\omega_s$), lower intensity
signal. Energy conservation is fulfilled through the generation of a
third beam (``idler'') whose frequency is constrained by \be \omega_p
= \omega_s + \omega_i. \ee Parametric gain is achieved over the
coherence length, defined as the length over which the phase
relationship among the three waves departs from the ideal condition
(``phase matching''). Phase matching corresponds to momentum
conservation and can be expressed as \be \vec{k}_p = \vec{k}_s +
\vec{k}_i, \ee $ \vec{k}_p,\ \vec{k}_s,\ \vec{k}_i$ being the wave
vectors of pump, signal and idler, respectively. Clearly, energy and
momentum conservation cannot be fulfilled simultaneously in a linear
crystal but birefringence offers a way out. In spite of the variety of
nonlinear crystals developed so far for frequency multiplication, only
a few can be grown to large size (tens of cm) to handle the pump
energy available and offer an adequate fluence limit for high power
applications. Here, we mainly concentrate on Potassium Dihydrogen
Phosphate (KDP), a negative uniaxial crystal commonly used for
frequency multiplication of Nd:YAG lasers\footnote{In the rest of the
paper we assume for KDP the following Sellmeier's equations: \be n_0^2
= 2.259276+ \frac{0.01008956}{\lambda^2 - 0.012942625} + 13.00522 \
\frac{\lambda^2}{\lambda^2-400} \ee for the ordinary index and \be
n_e^2 = 2.132668 + \frac{0.008637494}{\lambda^2 - 0.012281043} +
3.2279924 \ \frac{\lambda^2}{\lambda^2-400} \ee for the principal
extraordinary index ($\lambda$ is the wavelength in $\mu$m).  The
nonlinear coefficients are $d_{36} \simeq d_{14} = 0.44$~pm/V.}.  In
this case, phase matching can be achieved for parallel beams
(``collinear geometry'')
when the pump beam is at an angle $\theta_m$ with respect
to the KDP optical axis: \be n_{ep}(\theta_m) \omega_p \ = \ n_{os}
\omega_s + n_{oi} \omega_i. \ee Note that in the present configuration
the pump beam is polarized along the extraordinary direction, while
both the signal and the idler beam have ordinary polarization (``Type
I'' phase matching). Recalling \be n^{-2}_{ep}(\theta_m) = \sin^2
(\theta_m) n^{-2}_{ep} + \cos^2 (\theta_m) n^{-2}_{op} \ee $n_{ep}$
and $n_{op}$ being the principal extraordinary and ordinary refractive
indexes at pump wavelength, we get: \be \theta_m = \mathrm{asin}
\left[ \frac{n_{ep}}{n_{ep}(\theta_m)} \ \sqrt{
\frac{n^{2}_{op}-n^{2}_{ep}(\theta_m)}{n^{2}_{op}-n^{2}_{ep}} }
\right] . \ee It is worth mentioning that $\theta_m$ shows a less
pronounced dependence on the wavelength for Type I phase matching than
for type II, i.e. the case when only the idler or the signal has
ordinary polarization~\cite{ref:cerullo}. This is an additional
advantage when broad amplification bandwidth is sought for.

\begin{figure}
\centering \includegraphics[width=10cm]{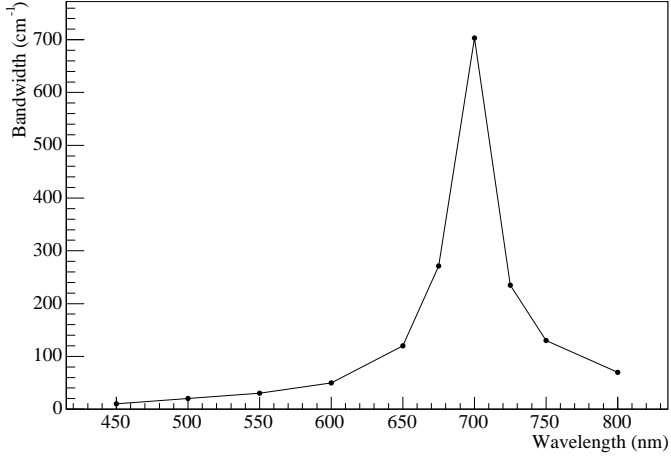}
\caption{FWHM bandwidth expressed in wavenumbers ($1/\lambda$) versus
signal wavelength for $\lambda_p=349$~nm. The KDP-based amplifier
(G=1000 at central wavelength) is operated in Type I collinear mode
(see text for details).}
\label{fig:band_vs_lambda}
\end{figure}

Fig.\ref{fig:band_vs_lambda} shows the FWHM amplification bandwidth
for a KDP-based Type I amplifier operated in collinear geometry.  The
bandwidth has been computed assuming a pump wavelength of 349~nm (see
Sec.\ref{sec:drivers}) and a pump intensity of 2~GW/cm$^2$.  The
latter is determined by the fluence F at which the crystal is operated
and the pump pulse duration $\tau$. Following~\cite{ref:ross}, we
assumed here F=1.0~J/cm$^2$ for KDP and $\tau$=0.5~ns~\footnote{In
fact, KDP can be operated at higher fluencies since its optical damage
threshold is greater than 5~GW/cm$^2$ but no long-term reliability
studies are available for these extreme values. Note also that
competing nonlinear processes like self-focusing or self-phase
modulations have been neglected.}. The gain bandwidth has been
computed assuming no pump depletion so that the gain G can be
approximated as~\cite{ref:ross,ref:armstrong}
\be G \ = \ 1+(\Gamma L)^2 \ \left[ \frac{\mathrm{sinh} B}{B} \right]^2 \ee
where $B\equiv \left[ (\Gamma L)^2 - (\Delta k L/2)^2 \right]^{1/2}$;
$\Gamma$ represents the gain coefficient
\be
\Gamma \equiv 4 \pi d_{eff} \sqrt{ \frac{ I_p }{2 \epsilon_0 
\ n_{ep}(\theta_m) \ n_{os} \ n_{oi} \ c \ \lambda_s \ \lambda_i } };
\ee 

\noindent the quantity $L$ is the length of the crystal and $\Delta k
\equiv k_p-k_s-k_i$ is the phase mismatch among signal, idler and
pump.  Note that in collinear geometry, this quantity is scalar since
the wave vectors lay along the same axis.  $d_{eff}$ is the effective
nonlinear coefficient: for Type I phase matching in KDP \be d_{eff} =
- d_{14} \sin \theta \sin 2\phi \ee where $\theta$ is the angle
between the propagation vector and the optic axis and $\phi$ is the
azimuthal angle between the propagation vector and the $xz$
crystalline plane\footnote{For the axis notation
see~\cite{ref:boyd}.}. Hence, $\theta=\theta_m$ and $\phi$ can be chosen
to maximize $d_{eff}$ ($\phi=\pi/4$).  In
Fig.\ref{fig:band_vs_lambda}, as well as in Ref.~\cite{ref:ross}, $L$
has been equalized in order to attain $G=1000$. In particular, for
$\lambda_s=700$~nm, such a gain is reached at $L=2$~cm.

\begin{figure}
\centering \includegraphics[width=8cm]{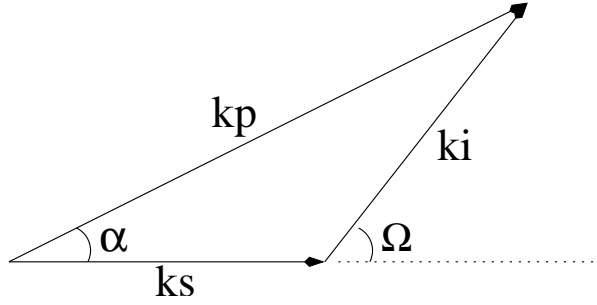}
\caption{Phase matching triangle for noncollinear OPCPA.}
\label{fig:vectors}
\end{figure}

Fig.\ref{fig:band_vs_lambda} points toward the existence of a window
for full exploitation of the original pump power. More precisely, a
faint, chirped, wideband seed signal could be amplified by a chain of
Type I amplifiers~\cite{ref:ross_opt_soc} and finally enter the power
KDP-based amplifier depleting the intense pump wave\footnote{A
complete numerical analysis of the signal evolution in pump depletion
mode is beyond the scope of this paper. A full calculation for
Nd:glass pumps has been carried out
in~\cite{ref:ross_opt_soc}.}. However, a significant improvement in
bandwidth can be achieved operating the system in non-collinear mode.
In this case the pump and signal wave vectors are no more parallel but
form an angle $\alpha$ between them (Fig.\ref{fig:vectors}). The
angle is independent of the signal wavelength. Again, the idler
frequency is fixed by energy conservation but the emission angle
$\Omega$ varies with $\lambda_s$. Therefore, the matching conditions
become
\begin{eqnarray}
\Delta k_{||} & = &  k_p \cos \alpha - k_s -k_i \cos \Omega = 0 \\ 
\Delta k_{\perp} & = &  k_p \sin \alpha -k_i \sin \Omega = 0 
\end{eqnarray}
The additional degree of freedom coming from the introduction of
$\alpha$ can be exploited to improve the gain bandwidth. In
particular, it helps achieving phase matching at first order for small
deviations from the central signal wavelength. 
It corresponds to imposing~\cite{ref:cerullo}
\begin{eqnarray}  
\left. \frac {\mathrm{d} \Delta  k_{||}} {\mathrm{d} \omega} 
\right|_{\omega=\omega_s} = 0 \label{eq:first_order1} \\
\left. \frac {\mathrm{d} \Delta  k_{\perp}}  {\mathrm{d} \omega} 
\right|_{\omega=\omega_s} = 0 
\label{eq:first_order2}
\end{eqnarray}
\noindent together with the energy conservation constraint (i.e. a
finite increase $\Delta \omega$ of the signal frequency corresponds to
a finite decrease $-\Delta \omega$ of the idler).

\begin{figure}
\centering \includegraphics[width=10cm]{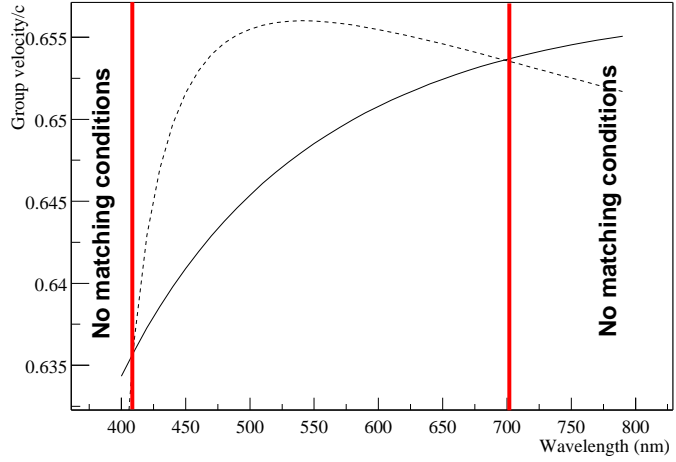}
\caption{(Continuous line) Signal group velocity as a function of
$\lambda_s=2\pi c/\omega_s$. (Dashed line) Idler velocity at $\omega_i
= \omega_p-\omega_s$ versus signal wavelength.}
\label{fig:groupvelocity}
\end{figure}

\noindent
Eqs.\ref{eq:first_order1} and \ref{eq:first_order2} are equivalent to:
\begin{eqnarray}
- \frac{ \mathrm{d} k_s} {\mathrm{d} \omega_s} + 
\frac{ \mathrm{d} k_i} {\mathrm{d} \omega_i} \cos \Omega - 
k_i \sin \Omega \frac{ \mathrm{d} \Omega} {\mathrm{d} \omega_i} & = & 0   \\
- \frac{ \mathrm{d} k_i} {\mathrm{d} \omega_i} \sin \Omega + 
k_i \cos \Omega \frac{ \mathrm{d} \Omega} {\mathrm{d} \omega_i} & = & 0   
\end{eqnarray}
and are simultaneously fulfilled if
\be
\frac{ \mathrm{d} k_i} {\mathrm{d} \omega_i} -  
\cos \Omega \frac{ \mathrm{d} k_s} {\mathrm{d} \omega_s} \ = \ 0.
\label{eq:match_noncoll}
\ee
The derivatives are related to the Sellmeier's equations for KDP
since 
\be
\frac{ \mathrm{d} k} {\mathrm{d} \omega} = \frac{n(\omega)}{c} +
\frac{\omega}{c} \frac{ \mathrm{d} n} {\mathrm{d} \omega}
\ee

\noindent
so that the signal/idler angle $\Omega$ can be explicitly
computed. These derivatives correspond to the group index for signal
($n_{gs} = c \mathrm{d} k_s/ \mathrm{d} \omega_s$) and idler ($n_{gs}
= c \mathrm{d} k_i/ \mathrm{d} \omega_i$). Hence,
Eq.\ref{eq:match_noncoll} can be interpreted as the request for signal
group velocity to equal the projection of idler group velocity along
the signal direction.  Note that it is impossible to fulfill
(\ref{eq:match_noncoll}) if the group velocity of the idler is smaller
than that of the signal.  For the case under consideration
($\lambda_p=349$~nm), this generalized matching condition can be
achieved in the signal region between 400 and
700~nm. Fig.\ref{fig:groupvelocity} shows the signal group velocity as
a function of $\lambda_s=2\pi c/\omega_s$ (continuous line). The
dashed line corresponds to the idler velocity at $\omega_i =
\omega_p-\omega_s$ versus the signal wavelength.  Finally, it is
possible to compute the (signal wavelength independent) angle $\alpha$
between the pump and the signal, which turns out to be

\be \sin \alpha = \frac {k_i}{k_p} \sin ( \mathrm{acos} \left[
n_{gi}/n_{gs} \right] ) \ee

\begin{figure}
\centering \includegraphics[width=10cm]{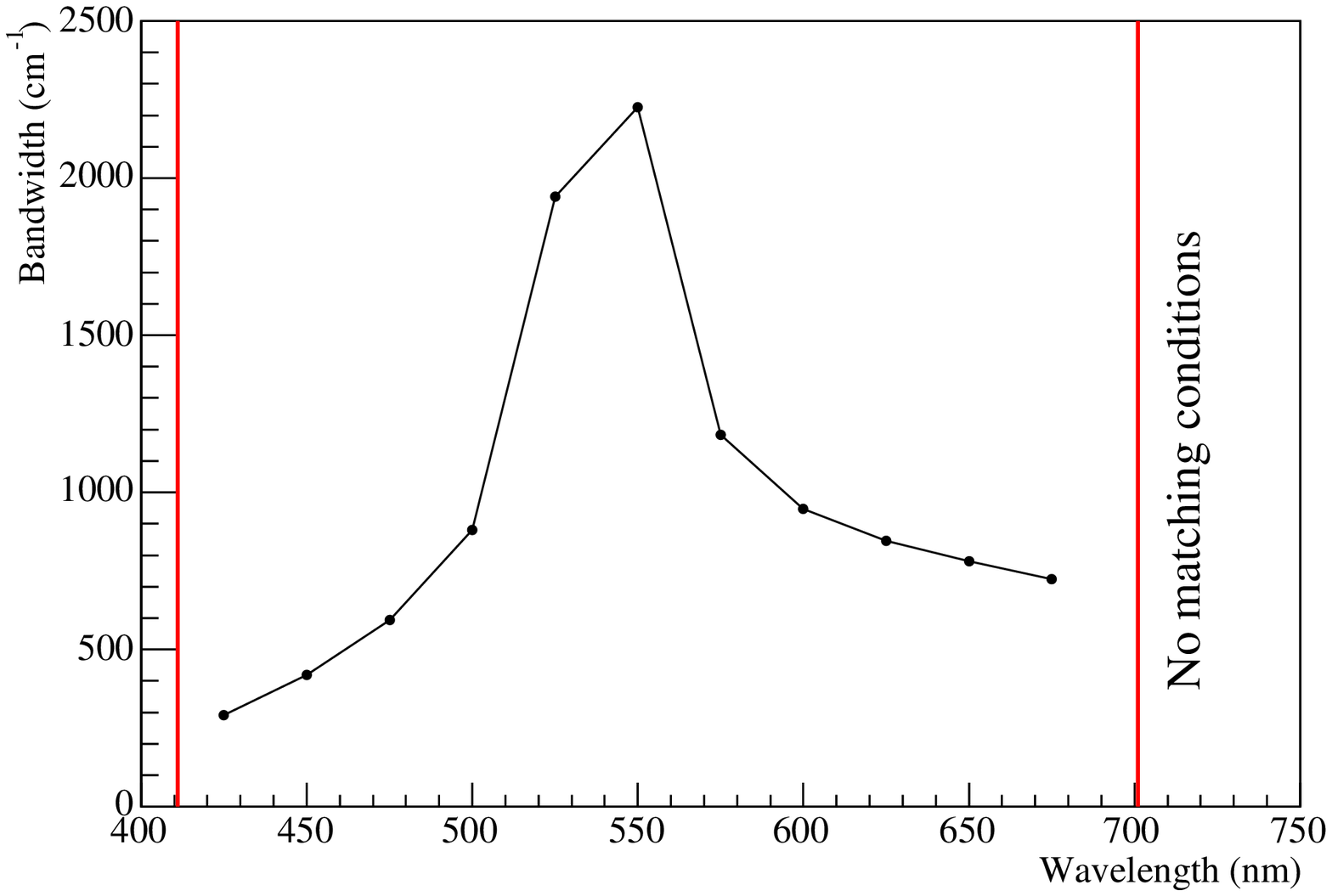}
\caption{FWHM bandwidth expressed in wavenumbers ($1/\lambda$) versus
signal wavelength for $\lambda_p=349$~nm. The KDP-based amplifier
(G=1000 at central wavelength) is operated in Type I non-collinear mode
(see text for details).}
\label{fig:band_vs_lambda_nocoll}
\end{figure}

\begin{figure}
\centering \includegraphics[width=10cm]{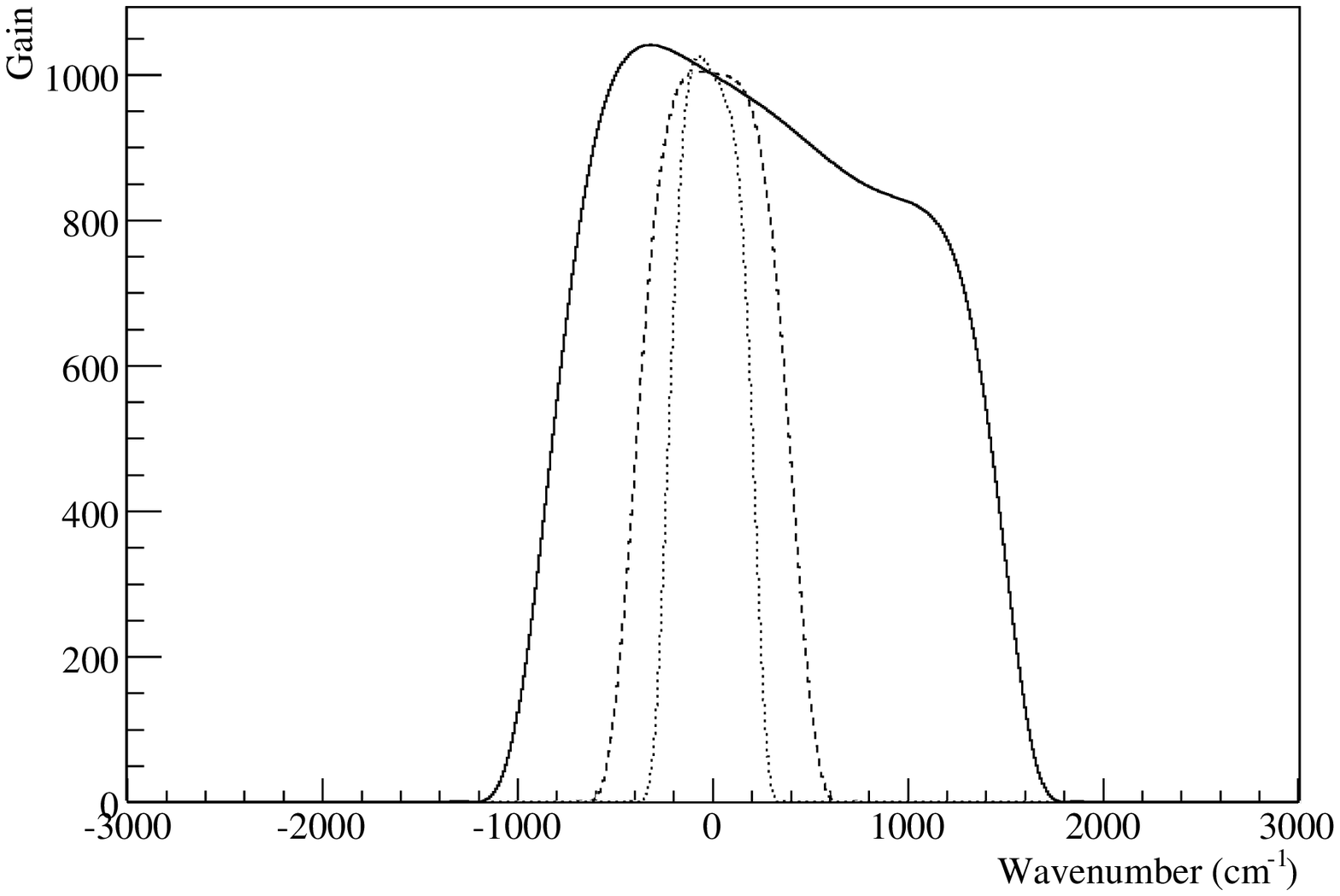}
\caption{Gain versus the wavenumber difference with respect to the
central wavenumber ($1/\lambda_s$), which the angle $\alpha$ has been
tuned for.  The continuous line refers to $\lambda_s=550$~nm (maximum
bandwidth).  The dotted and dashed lines refer to $\lambda_s=450$~nm
and $\lambda_s=650$~nm, respectively. The KDP-based amplifier (G=1000
at central wavelength) is operated in Type I non-collinear mode (see
text for details) at $\lambda_p=349$~nm.}
\label{fig:gain_vs_wavenumber}
\end{figure}

\noindent
The FWHM bandwidth of the amplified signal versus $\lambda_s$ for a
KDP-based Type I amplifier operated in non-collinear geometry is shown
in Fig.~\ref{fig:band_vs_lambda_nocoll}. As for
Fig.\ref{fig:band_vs_lambda}, the bandwidth has been computed assuming
a pump wavelength of 349~nm and a pump intensity of
2~GW/cm$^2$. Again, the crystal length $L$ for G=1000 is about
2~cm. Fig.\ref{fig:gain_vs_wavenumber} shows the variation of the gain
as a function of the wavenumber difference with respect to the central
wavenumber ($1/\lambda_s$), which the angle $\alpha$ has been tuned
for.  The continuous line refers to $\lambda_s=550$~nm (maximum
bandwidth).  The dotted and dashed lines refer to $\lambda_s=450$~nm
and $\lambda_s=650$~nm,
respectively. Figs.\ref{fig:band_vs_lambda_nocoll} and
\ref{fig:gain_vs_wavenumber} represent a key result: in non-collinear
geometry the broadband window for the exploitation of the DPSSL or XeF
drivers corresponds to a signal wavelength of about 550~nm. This
region is accessible by a modelocked Ti:sapphire oscillators (signal
generator) after self-phase modulation~\cite{ref:baltuska} and
frequency doubling. It provides the seed signal (at the nJ level) that
is amplified by the chain of low power amplifiers\footnote{See
e.g. Sec.3.1 of \cite{ref:ross}.}. The pump signal for the low power
amplifiers ($<$5~J) can be either derived by the main pump or, if
necessary, by a dedicated low energy pump at a more appropriate
wavelength. Finally, the signal is sent to the power amplifier
operating in pump depletion mode.  The studies performed in 2002 by
I.~Ross and coauthors~\cite{ref:ross_opt_soc} indicate that
extraction efficiencies of the order of 40\% can be obtained.  In
particular, a 4~kJ pump pulse would provide a broadband amplified
signal of about 1.6~kJ. The actual light on target depends on the
quality of the amplified signal and the compression optics and it is
discussed in the next section.

\section{Proton production and acceleration}
\label{sec:protons}

\begin{figure}
\centering \includegraphics[width=10cm]{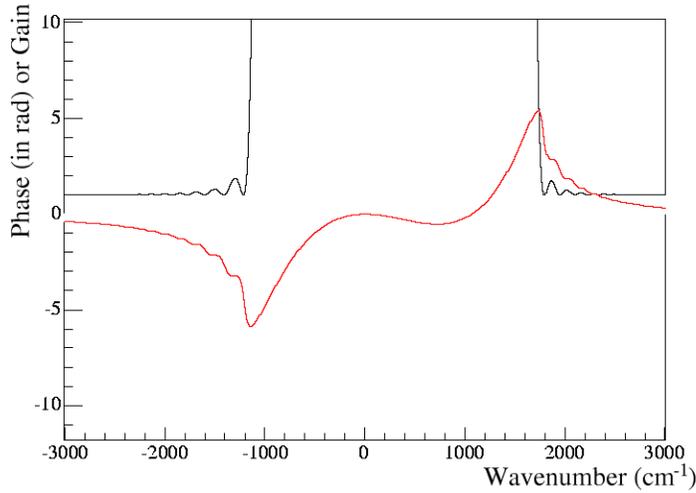}
\caption{Phase of the amplified signal (light line) and gain (dark
line) for the amplifier parameters of Fig.\ref{fig:gain_vs_wavenumber}
and $\lambda_s=550$~nm.}
\label{fig:phase}
\end{figure}

The pulse duration that can be achieved after the amplification
process is dominated, to first order, by the bandwidth of the seed
signal and the gain bandwidth of the OPCPA even if additional effects
connected to the beam quality entering the compressor and the
compressor itself should be taken into account. In particular, the
spectral phase~\cite{ref:armstrong,ref:ross_opt_soc} generated in an
optical parametric amplification when the seed signal is chirped plays
a role in setting up the recompression system.  For the case under
study, the phase $\Phi$ of the amplified signal is
given~\cite{ref:armstrong} by
\begin{equation}
\Phi = \mathrm{atan} \left[ \frac{ B \sin{(\Delta k/2L)} \ \mathrm{cosh}
B-(\Delta k/2L) \cos{B} \ \mathrm{sinh} B } { B \cos (\Delta k/2L) \
\mathrm{cosh} B + (\Delta k/2L) \sin (\Delta k/2L) \ \mathrm{sinh} B  } \right]
\end{equation}
and it is shown in Fig.\ref{fig:phase} (light line) for the amplifier
parameters of Fig.\ref{fig:gain_vs_wavenumber} and
$\lambda_s=550$~nm. For sake of clarity, the region where gain is $>1$ 
(dark line) is also shown. 

For this class of spectral chirping~\cite{ref:ross_opt_soc}, nearly
ideal recompression can be achieved as far as cubic phase terms can be
compensated. 
Neglecting the throughput efficiency of the compressor and the losses
due to the spectral clipping on the gratings, the output power $P$ for a
Gaussian profile is:
\be
P \simeq 1.6 \ \mathrm{kJ} \ \frac{\Delta \nu_{FWHM} \ \mathrm{(Hz)} }{0.44}  
\ = \ 240~PW
\label{equ:power_beamlet}
\ee
\noindent if the bandwidth is dominated by the OPCPA gain bandwidth at
$\lambda$=550 nm (2200 cm$^{-1}$).  The actual maximum intensity on
target depends on the quality of the optics and the available
compressor gratings. Note, however, that operating near diffraction
limit is not requested in the present case. In has been
shown~\cite{ref:esirkepov,ref:NIM_neutrino} that RPD acceleration
mechanism is fully operative for $I= 1 \times 10^{23} \
\mathrm{W/cm}^2$, although the transition region between the low
intensity regimes and the RPD one is presently unexplored both from
the experimental and from the numerical point of view. In the RPD
case, the energy of the accelerated ions depends 
on the intensity and pulse duration according to Eq. (17) in
Ref. ~\cite{ref:NIM_neutrino}
\begin{equation}
{\cal E}_{p,kin}=m_pc^2 \frac{2 w^2}{2 w+1}
\end{equation}
while the number of accelerated ions
depends solely on the illuminated area $S$~\cite{ref:NIM_neutrino}:
$N_p=n_0 S l_0$. Here $w$ is proportional to the laser pulse energy, ${\cal E}_{tot}$.
It is given by $w={\cal E}_{tot}/n_0 l_0 m_p c S$. The efficiency of the laser energy 
transformation into the fast proton energy is 
\begin{equation}
\eta=\frac{N_p{\cal E}_{p,kin} }{{\cal E}_{tot}}=\frac{2 w}{2w+1}.
\end{equation}

In the ultrarelativistic limit, $w\gg 1$,the efficiency tends to unity, 
and it is small in the nonrelativistic case when $w\ll 1$ . 

The studies performed in~\cite{ref:esirkepov} made use of 27~fs (FWHM)
gaussian pulses of $I=1.4 \times 10^{23}$~W/cm$^2$ and, at these
intensities, protons are accelerated following an $t^{1/3}$ asymptotic
law up to kinetic energies of $\sim 30$~GeV. These numerical studies
are extremely challenging even for large parallel computer facilities.
In order to reduce complexity, the study has been carried out with
laser pulse of relatively small focal spot. In addition, the dynamical
evolution has been followed up until the $t^{1/3}$ asymptotic
behaviour is reached (i.e. before the complete laser-plasma
decoupling).  The overall laser energy to proton kinetic energy
conversion efficiency at that time is 40\%. Extrapolation up to the
time of decoupling indicates that an energy conversion efficiency of
57\% can be reached and the maximum kinetic energy for the above
parameters exceeds 30~GeV. In the case under consideration, the
illuminated area corresponding to an intensity on target of $I=1.4
\times 10^{23}$~W/cm$^2$ is $S=1.5 \times 10^{-10} \ \mathrm{m}^2$,
i.e.  a circular spot of $7 \ \mu \mathrm{m}$ radius. Since the
kinetic energy reached by the protons is proportional to the product
of intensity and duration, a single shot (7~fs FWHM) would accelerate
particles up to about 8~GeV. The number of accelerated protons $N_p$
corresponds to the energy on target ${\cal E}_{tot}$ corrected for the
laser-to-ion energy transfer efficiency ($\eta$) divided by the
proton kinetic energy (${\cal E}_{p,kin}$).

\be N_p = \frac{\eta \ {\cal E}_{tot} }{ {\cal E}_{p,kin} } \ \simeq
\ 7\times 10^{11} \ \mathrm{protons/pulse} 
\label{eq:N_p}
\ee

\noindent
There is, however, an important caveat to be stressed. The RPD
mechanism is fully operative if the illuminated area is sufficiently
large so that border effects can be
neglected~\cite{ref:NIM_neutrino}. Due to limited computational
resource, a systematic study of the size and scaling of the border
effect is not available at present. Results from~\cite{ref:esirkepov}
indicates that, at the intensities mentioned above, the border corona
has a depth of the order of $5\ \mu\mathrm{m}$, This implies that
either the beamlet arrangement should be able to illuminate uniformly
a relatively large area, as it is done e.g. for the fuel target when
the driver works in ICF mode, or that a significant higher energy
should be used for the pump in a proper chain of KDP amplifiers. This
energy is not used to increase the pump or the signal intensity but
only the surface $S$, therefore it does not challenge the damage
threshold of the amplifier/compressor components.

\begin{figure}
\centering \includegraphics[width=12cm]{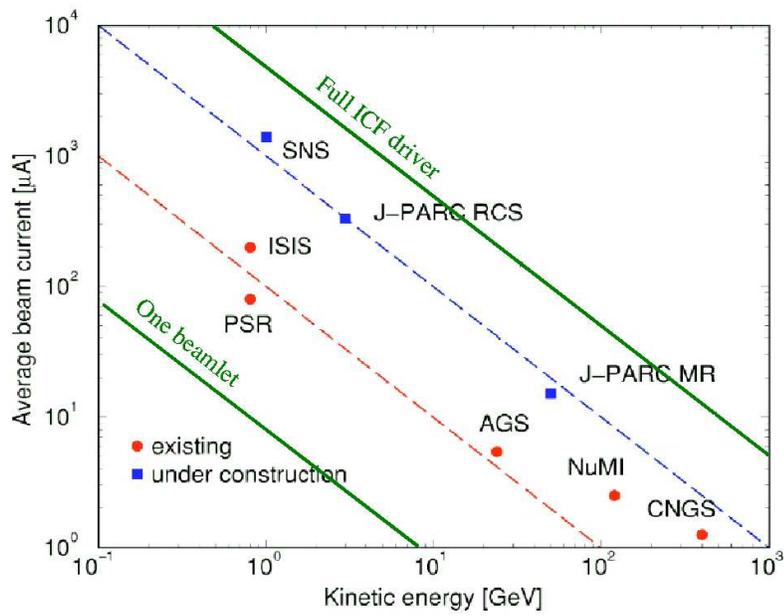}
\caption{Average proton current versus proton kinetic energy
for various existing facilities. The two green lines correspond
to a driver obtained by a single 4~kJ beamlet and by the full ICF facility
(2 MJ at 10 Hz rap.rate).}
\label{fig:list_drivers}
\end{figure}

From Eq.\ref{eq:N_p}, it follows that a single ICF beamline operating
at 10 Hz would be equivalent to a 10~kW proton driver. Clearly, a full
ICF facility (2 MJ pump energy at a repetition rate of 10~Hz) would
allow the construction of an extremely ambitious proton driver in the
multi-MW intensity range (see Fig.\ref{fig:list_drivers}). Differently
from traditional proton drivers, a laser driven device is highly
modular, while for RF-based accelerators particles have to be stacked
in a single lattice; in the RPD regime the stability conditions during
acceleration are much less stringent than for traditional drivers
since acceleration occurs nearly instantaneously and not in long
periodical structures. Anyway, the possibility to develop laser fusion
in strict connection with particle acceleration represents {\it per
se} a fascinating research line.

\section{Conclusions}

If experimentally confirmed, the radiation-pressure dominated (RPD)
acceleration mechanism~\cite{ref:esirkepov} will offer a stable and
efficient operation regime for laser-driven proton boosters. In
particular, it has been emphasized~\cite{ref:NIM_neutrino} the
opportunity to develop these facilities in conjunction with projects
for inertial confinement fusion (ICF) and neutron spallation sources.
In this paper, we have shown for the first time that the pulse
compression requests to make operative the RPD regime do not fall in
contradiction with the power requests of an ICF driver and we
discussed explicitly one solution based on optical parametric chirped
pulse amplification (OPCPA). Compatibility regions have been
identified for OPCPA amplifiers based on Potassium Dihydrogen
Phosphate operated in non collinear mode. In this configuration,
bandwidths exceeding 2000 cm$^{-1}$ (FWHM) have been obtained.

\section*{Acknowledgments}

The authors would like to thank M.~Borghesi and P.~Migliozzi
for insightful discussions and suggestions.


\end{document}